# A Physics-Informed Multi-Source Domain Adaptation Framework for Label-Free Post-Earthquake Damage Assessment

**ABSTRACT**

Efficient and intelligent assessment of post-earthquake structural damage is critical for rapid disaster response. While data-driven approaches have shown promise, traditional supervised learning methods rely on extensive labeled datasets, which are often impractical to obtain for damaged structures. To address this limitation, we propose a physics-informed multi-source domain adaptation framework to predict post-earthquake structural damage for a target building without requiring damage labels. The multi-source domain integrates actual damage data and numerical modeling data from buildings similar to the target structure. The framework operates through three key steps. First, the similarity of key physics from each domain are analyzed to form a weight matrix, which enhances domain differentiation. Second, features from the multi-source and target domains are extracted and fed into a classifier and a discriminator. The classifier ensures that the features are damage-sensitive and accurately assign damage states, while the discriminator enforces that the features remain domain-invariant. Finally, the key parameters matrix is applied as weights during adversarial training to optimize the contribution of features from each source domain. The proposed framework provides a robust solution for assessing structural damage in scenarios where labeled data is scarce, significantly advancing the capabilities of post-earthquake damage evaluation.

**Keywords:** Physics-Informed Learning; Domain Adaptation; Damage Assessment

## INTRODUCTION

Timely and reliable quantification of earthquake-induced structural damage is fundamental to protecting human life, guiding emergency operations, and formulating recovery strategies [1]. Within hours of a major seismic event, civil authorities must identify buildings at risk of collapse, deploy rescue teams, and allocate temporary shelter and repair resources in an informed manner. For decades, these decisions have depended primarily on manual visual inspection [2,3]: professional engineers perform systematic, building-by-building surveys, recording cracks, residual drifts, and other distress indicators before assigning safety tags. Although this practice is well established, it is inherently slow, labor-intensive, and exposes inspectors to considerable

Yifeng Zhang, Xiao Liang, Texas A&M University, 400 Bizzell St., College Station, TX 77843, U.S.A.

danger, particularly when aftershocks may trigger further structural degradation. Subjective judgements also lead to inter-inspector variability, and the procedure is difficult to scale to large metropolitan areas with thousands of buildings requiring rapid appraisal.

Model-based numerical simulation provides a complementary pathway. Given sufficiently detailed as-built documentation, a nonlinear finite-element (FE) model updated with recorded ground motions can estimate inter-story drifts, plastic hinge rotations, and component stresses, thereby supporting objective and spatially resolved damage assessments [4]. In practice, however, reliable FE simulations depend on accurate material properties, boundary conditions, and ground motion records. These information is often incomplete or uncertain in the chaotic post-earthquake environment. Thousands of nonlinear analyses must be executed to cover the range of possible excitations and parameter variations, imposing a computational burden that limits real-time applicability. The unavoidable modelling uncertainties further complicate the interpretation of numerical predictions in urgent decision contexts.

These limitations have stimulated a third line of research: data-driven structural-health monitoring (SHM). Rather than relying exclusively on constitutive formulations, data-driven approaches learn the mapping from measured dynamic response to damage state directly from historical observations. When labelled datasets exist, supervised machine-learning (ML) models could achieve high classification accuracy, for example support vector machine (SVM) [5,6], convolutional neural network (CNN) [7,8]. After a damaging earthquake, however, the very labels that supervised models require are unavailable. Instead, generating those labels is the goal of the assessment. Moreover, a classifier trained on one structure frequently generalizes poorly to another because the distribution of features and damage state labels varies with geometry, mass distribution, and boundary conditions. This domain transfer problem is particularly severe in civil engineering, where each building is effectively unique.

Domain adaptation (DA) offers a statistically principled solution by learning knowledge that remains damage-sensitive while becoming domain-invariant. Early SHM applications employed subspace-alignment techniques, such as Transfer Component Analysis (TCA) and Correlation Alignment, which project source and target data into a common latent space by minimizing kernelized statistical distances, typically the maximum mean discrepancy [9]. These methods improve robustness under moderate distribution differences; nonetheless, they rely on careful kernel selection, cannot easily handle missing sensors or variable record lengths, and deteriorate when the excitation intensity range differs substantially between structures [10].

Recent advances introduce adversarial DA [11], wherein a feature extractor is trained to deceive a domain discriminator while maintaining high classification accuracy—a minimax goal that encourages stronger distribution alignment [12,13]. Adversarial DA has enhanced performance in vibration-based damage localization, yet three critical challenges hinder its broader application for post-earthquake assessment. First, earthquake monitoring produces long, multi-channel time series acceleration sampled at high frequency. Extracting compact, informative features from such data is non-trivial [14,15]. Second, precise and specific frameworks require story-by-story damage estimates, not merely a global building state. Third, practical scenarios often provide labelled data from several buildings that differ in height, stiffness, or construction quality; balancing their influence is essential to avoid negative transfer.

To address these issues, we propose a physics-informed multi-source domain-adaptation (PI-MSDA) framework tailored to fully label-free post-earthquake damage assessment. The framework operates in three stages. (i) Physics-guided weight acquisition: relative height is used to form a similarity matrix that assigns weights to each labelled source domain, thereby favoring structures most similar to the unlabeled target domain. (ii) Adversarial feature learning: an one-dimensional convolutional neural network followed by a bidirectional long-short-term-memory (1D-CNN + Bi-LSTM) encoder processes raw acceleration records from all domains. The resulting embeddings feed a damage classifier, trained solely on labelled sources, and three domain discriminators, each distinguishing one source from the target. This joint objective forces the extracted features to remain damage-sensitive while becoming domain-invariant. (iii) The weights are assigned to each source domain and further loss calculation is performed. The weighted loss from predictor and discriminator is then transferred back to the feature extractor to perform further feature extraction.

The proposed methodology is evaluated using a comprehensive simulation campaign comprising nonlinear time-history analyses of 3 and 5-story reinforced-concrete frames subjected to 5,400 recorded ground motions. All response–label pairs from the 3-story building constitute the labelled sources, whereas the first story response from the 5-story building serves as the unlabeled target. Despite never accessing target labels during training process, PI-MSDA achieves high story-level accuracy, demonstrating its capability to support rapid screening under acute data scarcity.

## METHODOLOGY

The proposed multi-source adversarial domain-adaptation network is designed to transfer damage knowledge from labelled buildings to unlabeled building (target). The framework is organized into three blocks: feature extraction, damage prediction and discrimination. Together, these blocks create an system that learns damage-sensitive yet domain-invariant features and consequently delivers label-free damage estimates for the target building. The flowchart of this methodology is shown in Figure 1.

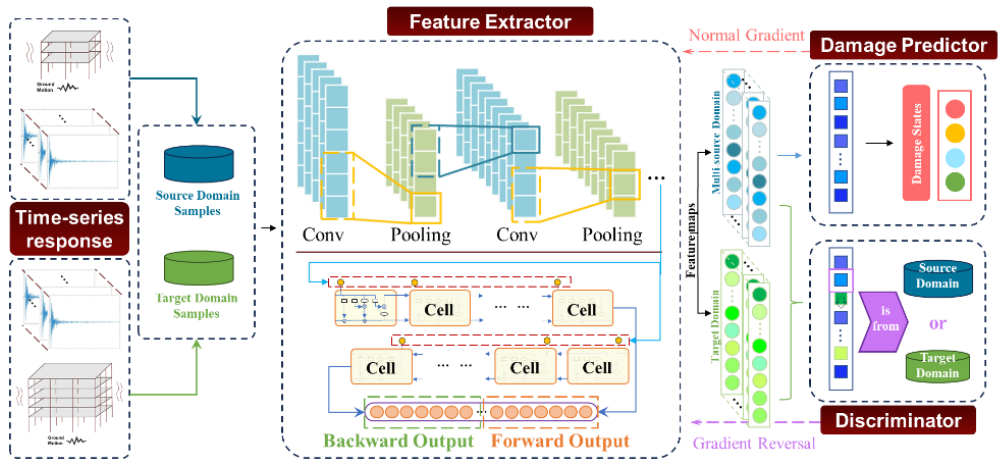

Figure 1. Framework of PI-MSDA.

**Source Weights Acquisition**

The weights are determined with the aim of classifying the degree of contribution of different source domains in the training process. The similarity between the physics is chosen as a measure of the degree of contribution of the different source domains. To achieve this, a Gaussian kernel is used to calculate the similarity between the different source domains as well as the target domain and this similarity value is used as the weights. The Gaussian kernel is calculated using the following formula:

$$S = e^{-\frac{(s-t)^2}{2\sigma^2}} \tag{1}$$

where, $s\ and\ t\ is\ the\ physics\ from\ source\ and\ target\ domain$, σ is the standard deviation of the $physics$ in source domain.

**Feature Extractor**

Denote by $X \in R^{L \times 4}$, a four-channel acceleration record of length $L$ obtained from a single story. The feature extractor maps the raw time series to a latent vector $z$ of dimension $d$. Its first stages are one-dimensional convolutions that act as data-driven filter banks, capturing local frequency content and reducing sequence length. The final stage is a bidirectional recurrent layer that aggregates long-range temporal dependencies. By design, $G_\theta$ is shared by every domain, consequently the same feature is imposed on source and target data.

$$z \in G_\theta(x) \tag{2}$$

$$z \in R^d \tag{3}$$

**Damage Predictor**

The classifier consists of one hidden layer and a soft-max output. It converts a latent vector into the posterior probabilities $\hat{y} = (p_1,\ p_2,\ p_3)$ of the three drift-ratio classes. Parameters $\phi$ are learned only from labelled source data; no target labels are used at this stage.

$$\hat{y} \in C_\theta(z) \tag{4}$$

$$\hat{y} \in R^3 \tag{5}$$

**Domain Discriminator**

Because three separate labelled source stories are available, a distinct discriminator is assigned to each source domain $S_i$. For discriminator $i$ the positive class corresponds to $S_i$ and the negative class to the target story $T$.

$$d_i \in D_{\psi_i}(z) \tag{6}$$

$$d_i \in (0,1) \tag{7}$$

Each $D_{\psi_i}$ is connected to $G_\theta$ through a gradient-reversal layer: during back-propagation the gradient entering $G_\theta$ is multiplied by $-\lambda$. The reversal causes $G_\theta$ to learn features that hinder the discriminators, thereby forcing the extracted features to become domain-invariant.

**Adversarial Training Routine**

During every training step a mini-batch containing samples from all sources and the target is assembled and forwarded according to

$$x \to G_\theta \to \begin{cases} C_\theta (damage\ prediction\ on\ sources) \\ D_{\psi_i}(domain\ discrimination\ for\ each\ Si) \end{cases} \quad (8)$$

The classifier receives supervision from source labels, whereas the discriminators learn to separate $Si$ from $T$. Because $G_\theta$ contributes to both tasks, it must balance damage sensibility against domain invariance.
The weighted classification loss is:

$$L_{clv} = \sum_{i=1}^{3} wi E_{(x,y)\in Si}[-y^T log C_\theta(G_\theta(x))] \quad (9)$$

where the inner term is the standard cross-entropy between one-hot label $y$ and prediction $\hat{y}$.
The adversarial loss for domain alignment is:

$$L_{adv} = \sum_{i=1}^{3} wi(E_{x\in Si}[log C_\theta(G_\theta(x))] + E_{x\in T}[log(1 - D_{\psi_i}(G_\theta(x))]) \quad (10)$$

a sum of binary cross-entropies that encourages each discriminator to distinguish its own source from the target.
The complete minimax objective couples both terms:

$$\min_{\theta,\varphi} \max_{\psi 1:3} (L_{clv} - \lambda L_{adv}) \quad (11)$$

For fixed weights $w$ and coefficient $\lambda$ the discriminators increase the adversarial loss, whereas the extractor and classifier decrease the total objective, thereby resulting in a desired compromise.

## CASE STUDY

**Building and Seismic Properties**

The damage-diagnosis framework is demonstrated on two three-dimensional reinforced-concrete special moment frames designed in accordance with current U.S. practice: a three-story structure and a five-story counterpart. Both buildings comprise two bays in each horizontal direction with uniform 24-ft spans and maintain a constant story height of 14 ft. Concrete and reinforcing steel are described by the *Concrete02* and *Steel02* constitutive models, respectively, and fiber-based force elements model all beams and columns. Each floor slab is assumed to act as a rigid diaphragm so that lateral response is governed solely by the frame action.

A M-7 seismic scenario is adopted for both frames. Horizontal ground-motion pairs are retrieved from the PEER NGA-West2 database. This query yields 180 ground motions. Each motion is subsequently assigned thirty scale factors, resulting 5 400 record–scale combinations are applied separately to each building, and nonlinear response histories are generated with the parallel version of OpenSees to carry out incremental dynamic analyses efficiently.

**Dataset Establishment**

The numerical simulation provides a complete record of floor-level responses and corresponding inter-story deformations. These outputs form the basis of the learning dataset used to train and evaluate the proposed domain-adaptation framework.
1. Data acquisition—floor accelerations
For the 3-story building, the nonlinear analysis yields a vector time series:

$$a^k(t) = [a_1^k(t), a_2^k(t), \ldots, a_n^k(t)] \quad (12)$$

for every ground-motion response k, where $a_i^k$ denotes the absolute horizontal acceleration at the i-th floor diaphragm. Each floor is treated as an independent observation. Further, for each floor, we use the time series acceleration of the floor and ceiling as a feature of this floor. Also, each acceleration contains both x and y directions. Therefore, the final data layout for each feature is as follows:

$$x = [a_{ix}^k(t), a_{iy}^k(t), a_{i+1x}^k(t), a_{i+1y}^k(t)] \quad (13)$$

2. Label generation—peak drift ratio
The same analysis provides the relative displacement histories $\Delta_i^k(t)$ between adjacent floors. The maximum value over the record length,

$$\delta_i^k = \max_t |\Delta_i^k(t)| \quad (14)$$

is normalized by the corresponding story height $h_i$ to obtain the peak drift ratio:

$$r_i^k = \delta_i^k / h_i \quad (15)$$

Each ratio is discretized into one of three damage states: Class 1 – $r_i^k \in (0,1\%)$; Class 2 – $r_i^k \in (1\%,2\%)$; Class 3 – $r_i^k \in (2\%,\sim)$. The resulting integer label set y aligns one-to-one with the feature samples in x.
Aggregating across all 5400 record per building, the procedure yields a balanced but diverse collection of floor-level acceleration features paired with categorical drift-ratio labels. The distribution of samples among the three classes for both the three-story and five-story frames is summarized in Table 1, providing a explicit view of class balance prior to domain adaptation.

TABLE I. THE LABEL DISTRIBUTION OF EACH CLASS

| Class | 1 | 2 | 3 |
|---|---|---|---|
| 5 Story Building | 38% | 40% | 22% |
| 3 Story Building | 39% | 41% | 20% |

3. Weight acquisition—relative height similarity

Relative height was chosen as the physical measure of similarity between the different domains. Although this physical information is too simple to represent the distribution of information between different domains, it can still help the model to learn the knowledge in different source domains better to some extent. Table 2 shows the physics and similarity of the source and target domains.

TABLE II. RELATIVE FLOOR HEIGHT SIMILARITY

| Story | 1 | 2 | 3 |
|---|---|---|---|
| 3 Story Building (source) | 1/3 | 2/3 | 3/3 |
| 5 Story Building (target) | 1/5 | / | / |
| Similarity | 97% | 69% | 33% |

**Result**

When the CNN+BiLSTM model is used directly to learn the source domain data, it can be found that the model tests well on the source domain, achieving an overall accuracy of 90%. However, when it is directly predicting the labels in the target domain, the performance is very poor and almost unclassifiable. The results are shown in Fig. 2, a) and b), respectively.

When using the proposed model, the predictive performance of the source domain shows some decline, with a reduced accuracy to 81%. However, the accuracy of the target domain showed a significant improvement to 72%. The results are shown in c) and d) in Fig. 2. This indicates that the knowledge of structural damage is transferred from the source domain to the target domain.

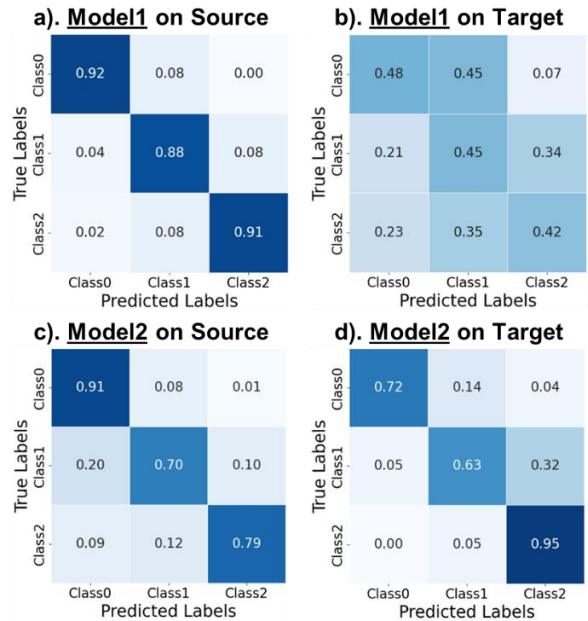

Figure 2. The performance of the prediction results (Model1: CNN+BiLSTM training only with source domain; Model2: Proposed framework).

In short, to some extent, these results suggest that the proposed domain-adaptation strategy can supply a rapid, label-free screening of post-earthquake drift demand.

**CONCLUSION**

Overall, the PI-MSDA predicts damage states with high accuracy on the labelled 3-story source building and, without any target labels, carries that knowledge to the first story of 5-story target building. While accuracy inevitably declines from the source to the target domain, the model still preserves a reliable ranking of damage severity on every level, identifying most damage cases correctly. These findings confirm that the proposed scheme, to some extent, can deliver a rapid, label-free first appraisal of floor-level drift demand, making it a practical aid for post-earthquake screening when detailed target labels are unavailable.